\documentclass{article}
\usepackage{frascatiphys_C}
\usepackage{graphicx}
\input ./babarsym
\def\Rdcs       {\ensuremath{R_{\rm D}}\xspace} % BaBar doesn't like 
\def\xPrimeSq   {\ensuremath{{x^{\prime}}^2}\xspace}
\def\yPrimeSq   {\ensuremath{{y^{\prime}}^2}\xspace}
\def\xPrime     {\ensuremath{x^{\prime}}\xspace}
\def\yPrime     {\ensuremath{y^{\prime}}\xspace}

\def\Dzbtokpi   {\ensuremath{\Dzb \to K^{+}\pi^{-}}\xspace}
\def\DztokpiWS  {\ensuremath{\Dz \to K^{+}\pi^{-}}\xspace}

\def\mKpi       {\ensuremath{m_{K\pi}}\xspace}

\def\dm         {\ensuremath{\delta m}\xspace}

\newcommand{\kevcc}{\ensuremath{{\mathrm{\,Ke\kern -0.1em V\!/}c^2}}\xspace}

\newcommand{\BABARPubYear}    {02}

\newcommand{\BABARProcNumber} {114}
\newcommand{\SLACPubNumber} {9552}

\begin{document}

\title{ 
Mixing in the \Dz-\Dzb system at \babar}
\author{
 Ulrik Egede\thanks{\hspace{0.2cm}
  On behalf of the \babar\ collaboration.}\\
\em Imperial College London, \\
    London SW7 2BW, \\
    United Kingdom.
}
\maketitle

\vspace{-14cm}
\begin{flushright}
SLAC-PUB-\SLACPubNumber \\
\babar-PROC-\BABARPubYear/\BABARProcNumber \\
%\babar-PUB-\BABARPubYear/\BABARPubNumber \\
%hep-ex/\LANLNumber \\
October, 2002 \\
\end{flushright}
\vspace{0.5cm}

\begin{center}
{\em Stanford Linear Accelerator Center, Stanford University, 
Stanford, CA 94309} \\ \vspace{0.1cm}\hrule\vspace{0.1cm}
Work supported in part by Department of Energy contract DE-AC03-76SF00515.
\end{center}

\vspace{8cm}

\baselineskip=11.6pt
\begin{abstract}
  We report a preliminary result for \Dz-\Dzb~mixing and the doubly
  Cabibbo suppressed decay rate~$\Rdcs$ based on an analysis of
  \DztokpiWS decays from 57.1~\invfb of data collected at or just
  below the \Y4S resonance with the \babar\ detector at the \pep2
  collider. We set 95\% confidence limits for the mixing parameters
  $\xPrimeSq$ and $\yPrime$ and find that our result is compatible
  with no mixing and no \CP violation. In the limit of no mixing we
  find the doubly Cabibbo suppressed decay rate $\Rdcs = ( 0.357 \pm
  0.022 \hbox{ (stat.)} \pm 0.027 \hbox{ (syst.)})\%$ and the
  $CP$~violating asymmetry $A_D = 0.095 \pm 0.061 \hbox{ (stat.)} \pm
  0.083 \hbox{ (syst.)}$.
\end{abstract}
\baselineskip=14pt

\section{Motivation}
\label{sec:Motivation}
Mixing can be characterised by the two parameters $x \equiv \Delta m /
\Gamma$ and $y \equiv \Delta\Gamma/2\Gamma$, where $\Delta m$
($\Delta\Gamma$) is the difference in mass (width) between the two
different mass eigenstates and $\Gamma$ is the average width.

Within the Standard Model the level of \Dz-\Dzb~mixing and \CP
violation is predicted to be below the sensitivity of current
experiments\cite{Petrov:2002qb}. For this reason it is a good place to
look for signals of physics beyond the Standard Model. Other
experiments\cite{Aitala:1998fg,Godang:1999yd} have already tried this
with smaller datasets using a technique similar to what is described
here. In any attempt to measure mixing one should consider the
possibility of \CP violation also as, with new physics, there is no
\emph{a priori} expectation that it is insignificant.

Mixing and \CP violation can be detected by observation of the
wrong-sign decay~\DztokpiWS (charge conjugation is implied unless
otherwise stated). Production through direct decay is doubly Cabibbo
suppressed~(DCS) but it is also possible for the \Dz to oscillate into
a \Dzb and subsequently decay through the right-sign Cabibbo
favoured~(CF) decay \Dzbtokpi. The two processes can only be
distinguished by an analysis of the time evolution of the decay.

Assuming $\xPrime$,~$\yPrime \ll 1$ and that \CP is
conserved, the time-dependent decay rate for the wrong-sign decay
$\Dz\ra\Kp\pim$ from DCS decays and mixing is
\begin{equation}
  \Gamma(\Dz\ra\Kp\pim)(t) \propto e^{-t/\tau_{\Dz}} 
  \left( 
    \Rdcs + 
    \sqrt{\Rdcs}\yPrime\; t/\tau_{\Dz} + 
    \frac{\xPrimeSq + \yPrimeSq}{4} (t/\tau_{\Dz})^2 
  \right)
  \label{eq:TimeEvolNoCPV}
\end{equation}
where $\tau_{\Dz}$ is the $\Dz$~lifetime and \Rdcs is the ratio of DCS
to CF decays\footnote{$\xPrime = x\cos\delta_{K\pi} + y
  \sin\delta_{K\pi}$ and $\yPrime = -x\sin\delta_{K\pi} + y
  \cos\delta_{K\pi}$ where $\delta_{K\pi}$ is an unknown strong
  phase.}. Because $\xPrime$ only appears in the time distribution as
a squared value, it is not possible to determine the sign of $\xPrime$
in an analysis based on the \DztokpiWS decay alone.

\CP violation can be either direct, in mixing or in the interference
between the two. The \CP violation gives rise to different apparent
values for the parameters in eq.~\ref{eq:TimeEvolNoCPV} so we define
$\Rdcs^{+(-)}$, $x^{\prime+(-)}$ and $y^{\prime+(-)}$ for $D$ mesons
produced as a \Dz~(\Dzb).

\section{The \babar\ detector and data selection}
\label{sec:babar}
For this analysis, we use $57.1~\fb^{-1}$ of data collected with the
the \babar\ detector which is described in detail
elsewhere\cite{Aubert:2001tu}. Reconstruction of charged particles and
particle identification of Kaons and pions are the most essential.
Tracking is provided by a five-layer silicon vertex tracker (SVT) and
a forty-layer drift chamber (DCH), both in a 1.5~T solenoidal magnetic
field. A Cherenkov ring imaging detector (DIRC) is placed outside the
tracking volume.

Kaons (pions) are identified by calculating a likelihood product of
the information from $dE/dx$ measurements in the SVT and DCH and the
reconstructed Cherenkov angle and photon statistics from the DIRC,
with an efficiency above 75\% (80\%) and mis-id rate below 8\% (7\%)
for $p < 4 \gevc$.

We select \Dz candidates from reconstructed $\Dstarp \ra \Dz\pip$
decays. The charge of the pion in the decay identifies the flavour of
the \Dz and also serves to create a clean sample of \Dz decays. Both
right-sign and wrong-sign \Dz candidates are selected. We select only
\Dstarp candidates with $p_{\Dstarp} > 2.6 \gevc$ in the
centre-of-mass frame to reject \Dstarp candidates from $B$ decays. Other
event selection criteria are employed to ensure that we have high
quality tracks and do not have any \Dstarp candidates with multiple overlapping
tracks.

\section{Results}
\label{sec:results}
\begin{figure}[htb]
  \centering
  \begin{minipage}[t]{0.48\linewidth}
    \includegraphics[width=\linewidth]{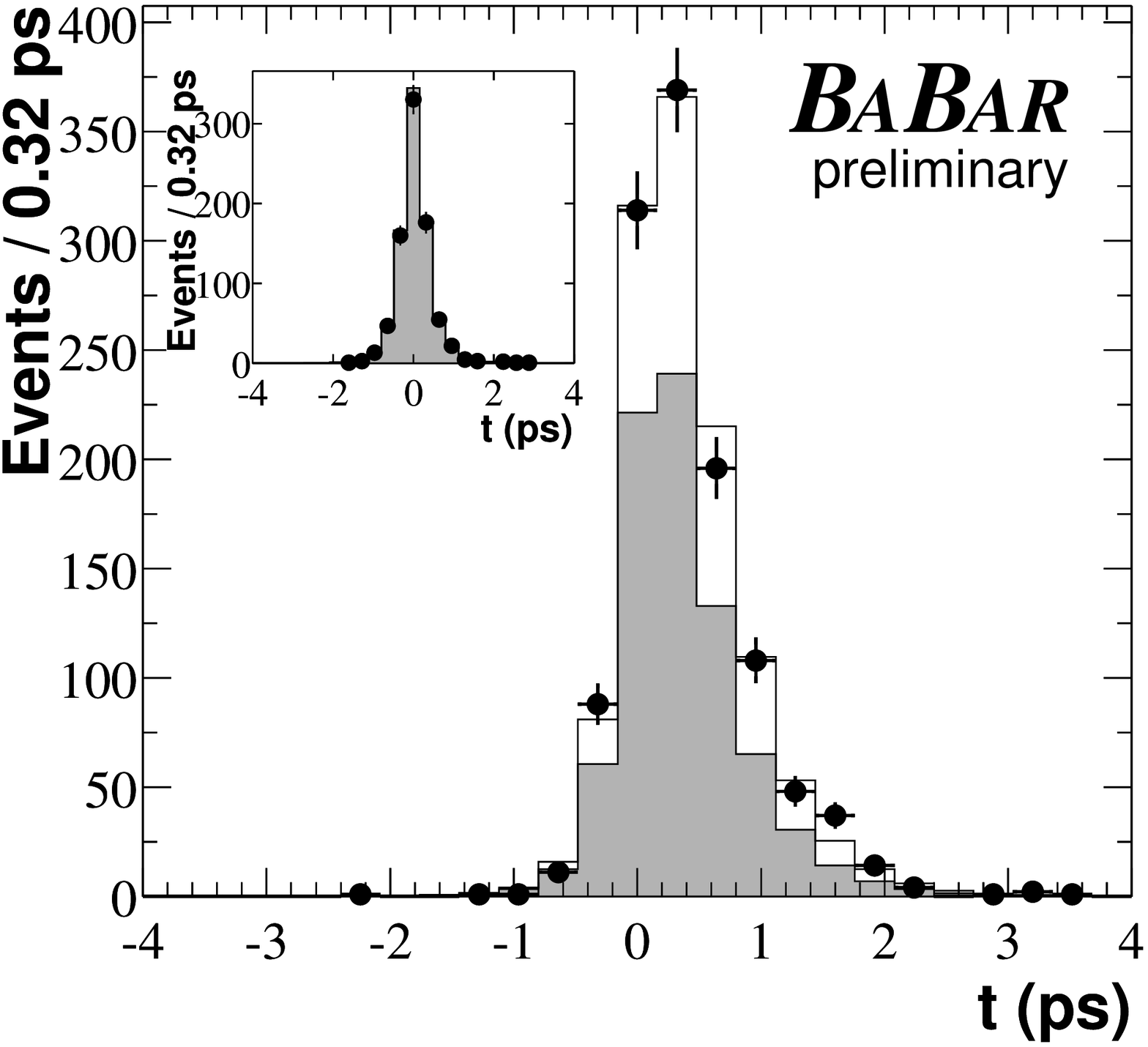}
    \caption{\emph{A projection of the wrong-sign data on the time
        axis with the main plot showing the signal region and the
        inset a sideband. The plots show data as points and the
        resulting projection of the background (shaded) and
        signal (open) from the fit.}}
    \label{fig:WStime}
  \end{minipage}
  \hspace{\fill}
  \begin{minipage}[t]{0.48\linewidth}
    \includegraphics[width=\linewidth]{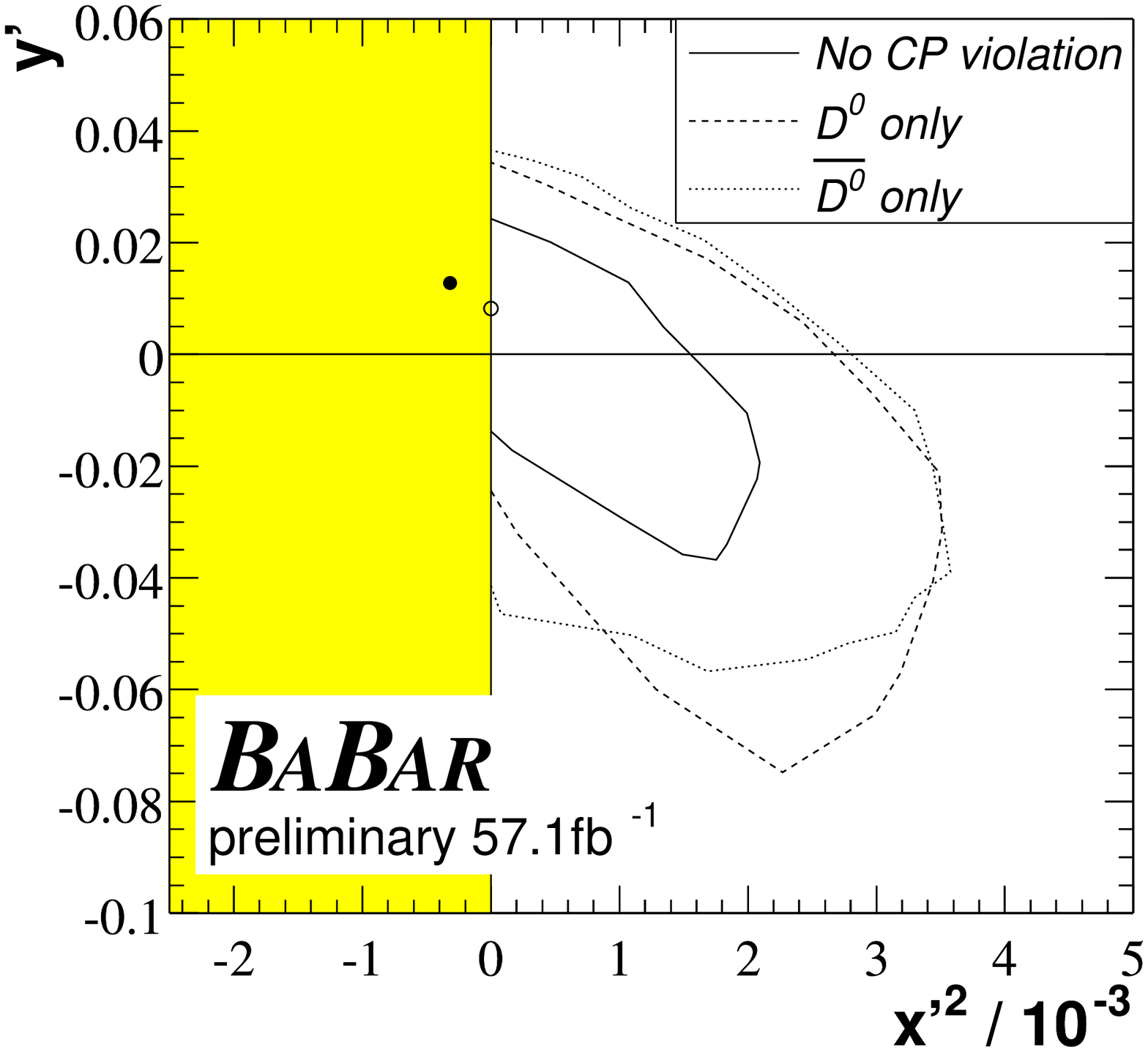}
    \caption{\emph{95\% confidence limits for \Dz (dashed) and
        \Dzb (dotted) separately (allowing for \CP violation) and the
        contour when no \CP violation is allowed (full). The solid
        point represents the most likely fit point in the case of no
        \CP violation and the open circle the same but forcing
        $\xPrimeSq>0$.}}
    \label{fig:Contours}
  \end{minipage}
\end{figure}

An unbinned maximum likelihood fit is used to extract the mixing
parameters. For each \Dstarp candidate we use the \Dz candidate
mass~\mKpi, the mass difference~\dm between the \Dstarp and the \Dz
candidate and the proper lifetime and error on the lifetime of the \Dz
candidate. The \Dz lifetime and the signal resolution model is
determined from the large right-sign sample. Sidebands are included in
\mKpi and \dm such that the level and time evolution of the different
background types can be evaluated. In figure~\ref{fig:WStime} we show
the time evolution of the wrong-sign sample. In total we observe
around 120,000 (440) right-sign (wrong-sign) signal events.

Since the fit allows $\xPrimeSq$ to take unphysical negative values an
error estimate from the log-likelihood surface~(LLS) would require a
Bayesian analysis where the choice of prior is not clear. In addition,
an accurate error estimate from the LLS requires a LLS shape that is
independent of the outcome of the fit. At the current level of
statistics these requirements are not even approximately met,
especially for the small mixing values observed.

Instead we use a method where we define a 95\% confidence limit
contour in $\xPrimeSq$ and $\yPrime$ space using toy Monte Carlo
experiments\footnote{With a toy Monte Carlo experiment we mean a Monte
  Carlo sample of the same size as the data generated from the PDF of
  the fit.}. Contours are constructed such that there is a 95\%
probability for any point $\vec\alpha_c=(\xPrimeSq_c, \yPrime_c)$ on
the contour that the likelihood ratio 
\begin{equation}
\Delta \ln {\cal L}(\vec\alpha_c)
= \ln {\cal L}(\vec\alpha_c) - \ln {\cal L}_{\rm max} \;,
\end{equation}
will be greater than the corresponding value $\Delta \ln {\cal L}_{\rm
  data}(\vec\alpha_c)$ calculated for the data. ${\cal L}_{\rm max}$
is here the maximum likelihood obtained from the fit to either data or
a toy Monte Carlo sample. The probability is evaluated by creating
multiple toy Monte Carlo samples at the point $\vec\alpha_c$ and for
each of the samples evaluate $\Delta \ln {\cal L}(\vec\alpha_c)$ after
a fit. 

As well as for the general case allowing for \CP violation we also
calculate our results for the special cases where \CP is conserved and
where no mixing is allowed.  In the case where we assume no mixing we
calculate the direct \CP violation term $A_D \equiv (\Rdcs^{+} -
\Rdcs^{-})/(\Rdcs^{+} +\Rdcs^{-})$.

\begin{table}[htb]
  \centering
  \caption{\emph{A summary of our results including systematic errors.
      A central value is reported for both the full fit allowing
      $\xPrimeSq<0$, and from a fit  with $\xPrimeSq$
      fixed at zero. The 95\% confidence limits are for the case where
      $\xPrimeSq$ is free.}}
  \vskip 0.1 in
  \begin{tabular}{lllll}
    \hline
     && \multicolumn{2}{c}{Fitted Central Value} &  \\
     Fit type & Parameter & $\xPrimeSq$ free & $\xPrimeSq$ fixed at 0 &
     95\% C.L. interval \\
    \hline
      & $\Rdcs^{+}$ [\%] & $0.32$   & $0.35$  & $0.18 < \Rdcs^+ < 0.62$ \\
      & $\Rdcs^{-}$ [\%] & $0.26$   & $0.27$  & $0.12 < \Rdcs^- < 0.56$  \\
      \CP violation
      & $x'^{+2}$  & $-0.0008$ & $0$  & $x'^{+2} < 0.0035$ \\
      allowed
      & $x'^{-2}$  & $-0.0002$ & $0$  & $x'^{-2} < 0.0036$ \\
      & $y'^{+}$ [\%]   & $\phantom{-}1.7$ & $0.7$ & $-7.5 < y'^{+}<3.4$ \\
      & $y'^{-}$ [\%]   & $\phantom{-}1.2$ & $0.9$ & $-5.7 < y'^{-}<3.6$ \\
    \hline
      No \CP
  &  $\Rdcs$ [\%] & $0.30$             & $0.31$  & $0.22 < \Rdcs < 0.46$ \\
      violation
  &  $\xPrimeSq$  & $-0.0003$          & $0$     & $\xPrimeSq < 0.0021$  \\
  &  $\yPrime$ [\%]    & $\phantom{-}1.3$ & $0.8$ & $-3.7 <\yPrime<2.4$ \\
    \hline
    No mixing 
  & \multicolumn{4}{l}{$\Rdcs = ( 0.357 \pm 0.022 \hbox{ (stat.)} 
      \pm 0.027 \hbox{ (syst.)})\%$} \\
  & \multicolumn{4}{l}{$A_D = 0.095 \pm 0.061 \hbox{ (stat.)} 
      \pm 0.083 \hbox{ (syst.)}$} \\
    \hline
  \end{tabular}
  \label{tab:resultsCPV}
\end{table}
The confidence contours for the mixing results including systematic
errors are shown in figure~\ref{fig:Contours} and the overall results
are summarised in table~\ref{tab:resultsCPV}.

For our systematic errors we evaluate the contributions from
uncertainties in the parametrisation of the PDF's, detector effects,
and effects of our selection criteria. For detector effects like
alignment errors or charge asymmetry we measure their effect on the
right-sign control sample. For variations in the event selection we
assign for this preliminary result the full variations in the
resulting contours as systematic errors.

In summary we have set new and improved limits on mixing and \CP
violation for neutral $D$ mesons. Our results are compatible with no
mixing and no \CP violation, all of which fits well with the
predictions from the Standard Model given our current sensitivity.

\bibliographystyle{h-physrev3}
\bibliography{frascatiphys_C}

\end{document}